# Flood Disasters and Health Among the Urban Poor

**Michelle S. Escobar Carías[1], David W. Johnston, Rachel Knott, Rohan Sweeney**

Centre for Health Economics, Monash University, Australia

Version April 20th, 2022

**Abstract**

Billions of people live in urban poverty, with many forced to reside in disaster-prone areas. Research suggests that such disasters harm child nutrition and increase adult morbidity. However, little is known about impacts on mental health, particularly of people living in slums. In this paper we estimate the effects of flood disasters on the mental and physical health of poor adults and children in urban Indonesia. Our data come from the Indonesia Family Life Survey and new surveys of informal settlement residents. We find that urban poor populations experience increases in acute morbidities and depressive symptoms following floods, that the negative mental health effects last longer, and that the urban wealthy show no health effects from flood exposure. Further analysis suggests that worse economic outcomes may be partly responsible. Overall, the results provide a more nuanced understanding of the morbidities experienced by populations most vulnerable to increased disaster occurrence.

---

[1] Corresponding author. Address: 900 Dandenong Road, Level 5, Building H, Caulfield Campus, Monash University, Caulfield East VIC 3145, Australia. E-mail: michelle.escobarcarias@monash.edu. Telephone: +61 0481-001030.

# 1. Introduction

Between 1990 and 2008, the share of urban poor in Asia rose by nearly 40%, and it is estimated that during the next 20 years, 95% of the population growth in developing countries will take place in urban areas (World Bank, 2020; Baker, 2012). This growing segment of urban poor faces a higher risk of exposure to natural disasters than both non-poor and rural populations (Winsemius et al., 2018; World Bank, 2020). Especially at risk are people living in slums and informal settlements[2], estimated at one billion people worldwide (UN-Habitat, 2016; World Bank, 2020). An absence of adequate urban planning has resulted in the development of these types of settlements in flood and cyclone-prone coastal areas, deemed undesirable for residential purposes (Baker, 2012). They are often characterized by dense and poorly constructed housing with low access to basic amenities and poor drainage (UN-Habitat 2016; CRED 2020).

When the urban poor experience floods, they face significant economic and health consequences. They may suffer damage to their homes, lose their livelihoods, and have increased exposure to injuries, pathogens and viruses (WHO 2018; Ezeh et al. 2017). Further, in the absence of timely and effective social safety nets, coping strategies employed in poor households include selling productive assets, reducing meals, or withdrawing children from school to enter the labor market. These can have long-term adverse effects on household income and consumption (Bui et al., 2014; Carter et al., 2007; Dercon, 2004), and human capital accumulation (Handa & King 2003; Carter & Maluccio 2003; Hoddinott & Kinsey 2001).

The risk of disastrous flood events in urban poor areas is increasing (IPCC 2007; UN-Habitat 2014), and so too is the need for high-quality evidence on how floods impact this vulnerable population. The literature on the health and well-being effects of high-impact low-frequency shocks in high-income country settings is growing. Economic studies have documented deleterious effects of large disasters such as severe floods, storms and hurricanes across the United States, Europe and Australia on birth outcomes (Currie & Rossin-Slater, 2013), disability-related measures of health among adults (Sastry & Gregory, 2013), post-shock mobility and long-run mortality (Deryugina & Molitor, 2020), mental illness (Baryshnikova & Pham, 2019), and on life satisfaction (Luechinger & Raschky, 2009). However, these findings are not necessarily applicable to low- and middle-income countries, where the majority of the world's urban-poor and informal settlement populations live. These populations are likely more vulnerable given their poor housing, inadequate sanitation and drainage, and unreliable water supply (Baker, 2012). Moreover, it is unclear whether the findings of these studies hold for settings that experience a high frequency of environmental shocks, where the cumulative effects of such events likely impede household and community resilience.

---

[2] UN-Habitat defines slum households as those that suffer from one or more of the following deprivations: lack of access to improved water source, improved sanitation, sufficient living area, housing durability and lack of security of tenure (UN-Habitat, 2003). The term informal settlement is usually used as synonymous but is defined by the UN according to three criteria: no security of tenure with modalities ranging from squatting to informal renting housing, lack of formal basic services and city infrastructure, and housing that does not comply with planning and building regulations often situated in environmentally hazardous areas. In addition, they can be occupied by urban residents of all income levels (United Nations, 2015).

In this paper we estimate the physical and mental health effects of flood exposure among the Indonesian urban poor, examining effects on both adults and children. Further, in recognition that residents of informal settlements may be especially vulnerable, we use two different datasets to separately explore the effects amongst the Indonesian urban poor living in formal housing and in informal settlements.[3] Indonesia is a country where urban residents experience regular environmental shocks (World Bank, 2019). This fact, combined with its large percentage of population living in slums (World Bank, 2016) and the emergence of high-quality longitudinal data, makes Indonesia a particularly informative setting for this research.

The relevant evidence from low- and middle-income countries is limited and largely based on post-disaster retrospective studies, with little focus on vulnerable populations.[4] One of the few exceptions is Datar et al. (2013), who show that for young children in rural India, a recent small to moderate natural disaster leads to an increase in acute illnesses such as diarrhea, fever and acute respiratory infections. Similarly, Rosales-Rueda (2018) finds that Ecuadorian children exposed to severe floods in utero were shorter in stature five and seven years later. In contrast to these two studies is Deuchert & Felfe (2015), whose results are suggestive of no substantial health effects. They find that super typhoon Mike in the Philippines had no impact on children's z-scores for weight and height, traditionally recognized as good indicators for short and long-term child health, respectively (Glewwe & Miguel, 2007).

---

[3] Only 15% of the urban IFLS sample fulfils more than one of the UN-Habitat (2003) characteristics for slum households. Therefore, overlap between both datasets is unlikely.

[4] A large group of studies, mostly in medical literatures, present cross-sectional analyses of data with retrospective reports of disaster occurrence. They mainly focus on adult samples that seek medical attention and explore short term effects. Results indicate substantial increases in the prevalence of gastrointestinal symptoms (Carlton, et al., 2014; Davies, et al., 2015; Ding, et al., 2013; Zhang, et al., 2019), cough and fever (Portner, 2010), skin and eye infections and injuries (Saulnier et al. 2018; Wu et al., 2015; Bich et al. 2011), and post-traumatic stress disorder (Norris et al., 2004).

One reason for the limited evidence from developing countries is the scarcity of longitudinal data that can be used to overcome the two main methodological challenges. First, while many disaster events are temporally random, they are not spatially random, which biases the simple comparison of affected and unaffected people. For example, low income households are often forced to reside in locations with low environmental quality and high risk of disaster (Winsemius et al., 2018). Second, if data collection occurs solely after the event, researchers are likely to encounter sampling challenges as disasters may scatter potential participants - sometimes the most affected ones - outside the affected area (Galea et al., 2008; Kessler et al., 2006).

To address these challenges, we use two longitudinal datasets – the Indonesia Family Life Survey (IFLS) (Strauss et al., 2016) and the Revitalising Informal Settlements and their Environment (RISE) survey (Leder et al., 2021). We first use the 2007-2008 and 2014-2015 waves of the IFLS to test the extent to which exposure to floods affects the physical health of adults and children, and mental health of adults, living below the poverty line. The IFLS, however, does not distinguish between the urban poor living in formal and informal settings, nor does it collect data on child mental health. We therefore extend the analysis with the RISE survey data, which collects measures of adult and child physical and mental health twice per year from a sample of informal settlements in Makassar, Indonesia.[5] The city of Makassar itself experienced a major flood event in January 2019 – one of the largest ever recorded in the province of South Sulawesi –, and the RISE survey collected data shortly before and afterwards. This allows for the observation of immediate flood impacts, helps rule out potential confounders such as selective mortality and migration, and allows measuring flood effects on child emotional functioning.

Fixed-effects analysis of the two data sets provide several important findings. According to our estimates, exposure to flooding and the subsequent housing damage has substantial negative impacts on physical health. This effect mostly dissipates within 12 months of the shock; however, the size of the effects four to five months post shock has serious implications for the health and living standards of the Indonesian urban poor. This is however not the case for individuals beyond the poverty threshold who, in contrast, report no significant changes in health status associated with flood exposure. Methodologically, we show that frequent sampling, as per the RISE survey, is important for identifying the immediate health effects caused by natural disasters. These insights are less likely to be captured by longitudinal surveys with years between waves.

---

[5] The RISE survey is conducted as part of an ongoing randomized controlled trial designed to test a "water-sensitive-cities approach" (Brown, et al., 2018) in 12 informal settlements in the city of Makassar.

The experience of floods in the past 12 months is also estimated to significantly worsen the mental health of both urban poor adults and children. In particular, children living in informal settlements experienced a substantial decrease in mental health, with a 78% worsening in their emotional functioning score, relative to the sample mean. We provide some evidence that the mental health effects may be partly driven by increased economic and financial stress. Flood disasters are estimated to significantly increase medical expenditures, decrease wealth, and increase attempted borrowing. These findings are robust to different ways of defining poverty status to delimit our sample of interest.

Overall, our findings contribute to the international literature investigating the links between environmental shocks and physical and mental health, and more generally to the emerging stem of economic research on life in slums and informal settlements (Binzel & Fehr, 2013; Bird et al., 2017; Marx et al., 2019; Alves, 2021). These findings are relevant beyond Indonesia. Millions of the world's urban poor experience environmental shocks of this type, and will continue to do so with increasing frequency due to climate change.

The rest of the paper proceeds as follows. Section 2 describes the Indonesian context and provides a brief overview of the mechanisms through which floods affect human health. Section 3 outlines the sources of data, outcome and treatment measures. Section 4 outlines the empirical strategy and findings for the IFLS sample. Section 5 provides a similar detailed account of the empirical specification and findings for the RISE sample. Section 6 discusses financial effects of floods, and section 7 concludes.

**2. Background**

*2.1 Floods in Indonesia*

Indonesia is the world's largest archipelago, and has approximately 247 million inhabitants. It is also the world's fourth most populous country. As a result of its geographical location on the "Ring of Fire" and its spread to the north and south of the Equator, it is exposed to a variety of disasters including droughts, earthquakes, floods, landslides, storms, volcanic activity, tsunamis and wildfires. Using data from EM-DAT[6], Figure 1a shows that Indonesia has had the third most flood events of any country between 1970 and August 2020, and Figure 1b suggests that the frequency of harmful flooding events is increasing over time. Possible reasons for this trend are climate change and the better reporting of hydrological phenomena.

---

[6] EM-DAT: The Emergency Events Database (Centre for Research on the Epidemiology of Disasters – CRED).

[Figure 1]

Floods were by far the most frequent natural disaster in the past 50 years in Indonesia. They affected more people than other natural disasters and were the second deadliest after earthquakes and tsunamis (CRED, 2020). A climate change vulnerability assessment of the city of Makassar, which includes our sample of informal settlements, concluded that although rainfall levels would remain constant in the coming years, precipitation would be concentrated over a shorter period of time. This implies a prolonged dry season and more intense rainfall causing flooding (UN-Habitat, 2014). Improving our understanding of the impacts of floods on the millions of Indonesians living in urban poverty is therefore clearly important.

*2.2 Floods and Health*

Floods are estimated to have caused more than 55,000 deaths globally over the last ten years (CRED, 2020), and are associated with water-, vector- and rodent-borne diseases. According to the WHO (2020), the risk of water-borne diseases increases when there is significant population displacement, when drinking water sources are compromised, and through direct contact with polluted waters; thus, disease outbreaks are more likely to happen in low-resource countries. Diarrheal infections and fever are two of the most commonly found water-borne diseases (Vollaard et al., 2004; Wade et al., 2004). Other common water-borne diseases include skin infections and upper respiratory infections caused by growth of mold following floods (Watson et al., 2007; Saulnier et al., 2018; Wu et al., 2015; Bich et al., 2011). Further, as flood waters recede, they can provide the ideal breeding sites for mosquitoes transmitting diseases such as malaria and dengue (WHO, 2020).

The medical literature also shows that severity of exposure, and preexisting mental illness, are key predictors for mental health problems following a natural disaster (Sullivan et al., 2013). For instance, Acierno et al. (2009) find that typhoon related consequences such as forced evacuation, injury due to storm and pre-exposure mental health status might be important predictors of negative psychological outcomes in Vietnam. Norris et al. (2004) find that communities which had experienced mass casualties and displacement after floods and mudslides in Mexico showed a higher prevalence of PTSD and major depressive disorder. Bereavement, loss of property and employment, as well as stress over lack of food and shelter have also been correlated with mental illness after disasters (Galea et al., 2007).

*2.3 Burden of Disease in Indonesia*

Despite substantial improvements in health insurance coverage amongst poorer individuals and families in the last decade (Mboi et al., 2018), lower respiratory infections, tuberculosis and diarrheal disease remain among the leading causes of disability-adjusted life-years in Indonesia. According to the latest DHS survey, 14 to 20% of Indonesian children under 5 experienced episodes of diarrhea in the past two weeks (BPS et al., 2018). Diarrheal disease is the third leading cause of death amongst Indonesian children, and the tenth leading cause of death across all age groups (Ayu et al., 2020). Approximately 43% of children and 37% of adults reported symptoms of respiratory infections in the fifth wave of the IFLS survey.

Mental health is another pressing issue in Indonesia. Between 2013 and 2018, the prevalence of depression increased from 3.7% to 6.1% (Prastyani, 2019), though mental health is highly stigmatized in Indonesia, so estimates likely underestimate the real burden (HRW, 2016). There is an apparent economic and educational gradient to emotional disorders, with a substantially higher prevalence amongst the lowest income quintile and adults with no schooling (Ministry of Health of Republic of Indonesia, 2013). Most Indonesians have very poor access to trained mental health specialists with one of the world's lowest psychiatrist-to-population ratios of 0.31 per 100,000 people and 0.18 psychologists per 100,000 people (WHO, 2019). The geographic distribution of this workforce is highly concentrated on the island of Java where about 70% of psychiatrists are based, 40% of whom work in the capital, Jakarta, where only 4% of the Indonesian population live (HRW, 2016).

**3. Data**

*3.1 The Indonesia Family Life Survey*

To assess the impact of floods on urban poor living in formal housing, we use a balanced panel from the last two waves of the Indonesia Family Life Survey (IFLS). The IFLS is a longitudinal study based on a sample of households in 1993 that reportedly represents 83% of the population in Indonesia from 13 provinces (Strauss et al., 2016). Four subsequent waves were collected in 1997 (IFLS2), 2000 (IFLS3), 2007-2008 (IFLS4) and 2014-2015 (IFLS5). Our estimation sample comprises households located in urban areas living below the poverty line in IFLS4 or IFLS5, where poverty is defined as having an equivalized consumption of less than $1.51 per

day (ADB, 2016).[7,8] These urban and expenditure restrictions left us with a balanced sample of 8,006 adults (age 15+) and 4,072 children (age 0-14). Summary statistics of the adult and child sample are presented in Table A1 in the Appendix.

Information on natural disaster occurrence was collected in IFLS4 and IFLS5. Respondents were asked whether there had been a natural disaster in their area of residence during the past five years, and if there had been, the type of disaster. In our sample, 89.7% of adults hadn't experienced a flood in either sample period, 3.1% experienced a flood between 2002-2008, 3.5% experienced a flood between 2010-2015, and 3.8% experienced a flood in both time periods. Respondents were subsequently asked the year when the largest flood occurred. Of the 10.3% of respondents who reported flood exposure, 16.8% reported that the flood had occurred in the past year, 7.1% reported that the flood occurred more than one year ago, and 76.1% could not recall. We assume that those who could not recall experienced the flood more than one year before the survey and are reclassified as such.

The health outcomes of interest in the IFLS are poor self-assessed health and the experience of acute morbidities (e.g. stomach ache, nausea, etc.) for both adults and children. For adults only, we also explore depression. These outcomes are described in Table 1. Around 22% of adults and 11% of children are somewhat or very unhealthy, and both adults and children experienced around two acute morbidities in the four weeks prior to the survey. Depression is measured by the CESD-10 score which represents the frequency of 10 depression-related feelings during the past week, and is scored on a scale of zero to 30, increasing in depressive symptoms (Andresen et al., 1994). A score of 10 or above indicates that a person may have depression. The IFLS sample has a mean of 7.8 and a standard deviation of 4.5, with 24% of people scoring above the 10-point threshold for potential depression.

[Table 1]

*3.2 Revitalising Informal Settlements and their Environment Survey*

The Revitalising Informal Settlements and their Environments (RISE) project is an action-research program testing a water-sensitive-cities approach to upgrading 12 informal settlements in Makassar, Indonesia, and 12 settlements in the Fijian capital city of Suva (Leder et al., 2021; French, et al., 2021).[9] We use data from three 6-monthly RISE household surveys completed in Indonesia between November 2018 and December 2019, prior to any upgrading of the settlements. Wave one took place in November-December 2018, wave two in May-July 2019, and wave three in November-December 2019. All households in the 12 Indonesian settlements were approached to be surveyed, with 90% of households completing the face-to-face survey. We have information on 500 adults (1,430 observations), the majority of which are females[10], and 580 children (1,715 observations). Descriptive statistics of characteristics measured at baseline are presented in Table A2.

---

[7] It is possible that household consumption was affected by flood exposure, and therefore be endogenous. However, we show in Section 4 that our estimates are similar when the sample is instead selected based on two alternative village-level poverty indicators measured in IFLS4.
[8] We tested whether respondents that left the study between IFLS4 and IFLS5 did so because of floods reported in IFLS4. These results are shown in Appendix Table A3. We find that flooding is not a significant predictor of death before IFLS5 nor is it a predictor of the decision to migrate or move out of the household.

The RISE survey data have several key features that allow for insights not possible with the IFLS. First and most importantly, the data allowed us to estimate how floods impact the impoverished residents of informal settlements (slums). Few data sets include respondents from informal settlements, and if they do, the sample sizes are usually small. Second, the RISE survey data include information on the mental health of children aged 5 to 14 years old. Child mental health was measured using the emotional well-being module of the parent-proxy Paediatric Quality of Life Inventory$^{TM}$ 4.0 Generic Core Scales (PedsQL) questionnaire (Varni, Seid, & Kurtin, 2001). The score ranges from zero to 100, with higher scores representing worse emotional functioning. Similar to the IFLS, adult mental health was measured using the CESD-10 depression score (waves one and three only). Physical health was measured using self-assessed general health and number of acute morbidities experienced in the past week. The health outcomes are also described in Table 1.

When compared against the statistics of the national urban poor sample in the IFLS, we see that the RISE sample is more likely to be in poor general health, 28% of adults and 17% of children reporting that their general health was very bad, bad or moderate. They report a lower number of acute symptoms of 0.53 for adults and 0.59 for children, but this partly reflects the different reference periods (one week in RISE, four weeks in IFLS). In terms of adult mental health, the mean CESD-10 score for the informal settlement residents is 6.5, which is lower than the IFLS sample. In regards to child emotional functioning, RISE children age under 8 have an average score of 82.55, and children aged 8 and above have an average score of 82.66. According to Huang, et al. (2009), for children under 8 years, the cutoff scores to identify potential special health care needs, moderate conditions, and major conditions are 83, 79 and 77, respectively. For children ≥ 8 years, the study recommends cutoff scores of 78, 76 and 70, respectively. According to these cutoffs, the score of the average RISE child under 8 is indicative of an emotional functioning problem. For practical uses, we reverse the emotional functioning scale such that for all outcome variables, a higher regression coefficient reflects a deterioration in either physical or mental health status. Thus, the reversed mean score reported in Table 1 is 17.38.

---

[9] RISE uses a parallel-cluster, randomised controlled trial design. Six settlements in each country were randomly assigned in 2019 to the intervention group and six to the control group, which will receive the upgrades at a later phase. The intervention is site-specific, co-designed with each community, and adapted to the needs in each informal settlement. See www.rise-program.org

[10] RISE household surveys mainly targeted female caregivers and female heads of households.

The third key feature of the RISE survey data are specific questions about a major flooding event that occurred in Makassar during the last week of January 2019. These were asked in the weeks immediately after the flood which reduces concerns regarding potential recall bias. This also enables us to estimate the health effects of a known flood event, rather than the effects of generic flood events (of varying severity levels) as we do with the IFLS. The specific questions we rely on were asked in wave two, approximately four to five months following the flood, and allow the construction of a variable indicating whether the flood event caused damage to the household's house, land or other assets. Almost 30% of our adult and child samples experienced this direct damage.

## 4. Health Effects of Floods on Urban Poor Respondents in the IFLS

We estimate the effects of flood exposure using regressions with *individual* and *area-year* fixed effects. Specifically, the health of individual $i$ residing in province $p$ in time $t$ ($hlth_{ipt}$) is regressed on a set of flood indicators ($flood_{ipt}$), time-varying control variables ($X_{ipt}$), individual fixed-effects ($\alpha_i$), and province-year fixed-effects ($\tau_{pt}$):

$$hlth_{ipt} = \delta flood_{ipt} + X'_{ipt}\beta + \alpha_i + \tau_{pt} + \varepsilon_{ipt} \tag{1}$$

where $\alpha_i$ are individual fixed-effects included to control for potentially omitted variables that are constant over time and which could affect individuals risk preferences, choice of residence, and other unobserved characteristics that can affect both flood exposure and health status. $X_{ipt}$ is a vector of individual and household characteristics, including a cubic function of respondent's age and household head's age, the number of children and number of adults in the household.[11] Province-year fixed-effects are also included, in addition to the individual fixed-effects, to control for all unobserved time-varying province-level determinants of health. We do not include village fixed-effects because flooding may have caused individuals to change their location of residence, and the data shows such moves are more likely to have occurred within provinces than across them.[12] Robustness checks presented in Table A5 using regency-year fixed effects, the next geographical unit below provinces, show that the results of our preferred specification in Table 2 are robust for both adults and children.

---

[11] Covariates representing household socioeconomic status (SES) are not included because they may be partly determined by flood occurrence. However, the inclusion of a more extensive set of (possibly endogenous) control variables has little effect on our reported estimates. The results of these regressions are reported in Table A4.

The main parameter of interest is $\delta$. It is identified by the 6.6% of individuals who reported a flood in one wave, but not the other. The main threats to causal identification are time-varying individual-level omitted variables that are associated with both flood exposure and health. A possible candidate is household economic status, which is likely to impact health outcomes and location of residence. We explore this possibility by using data from waves 3 to 5, and regressing future ($t+1$) flood exposure on current ($t$) socioeconomic circumstances, and individual and province fixed-effects. The results are reported in Appendix Table A6. They indicate that changes in a person's circumstances over time are not predictive of future flood exposure: F-statistic equals 0.61 ($p = 0.893$).

Table 2 presents estimates of the flood effects ($\delta$ in equation 1) separately for adults and children, and for three IFLS health outcomes: poor general health, number of acute morbidities, and depression score. Adults who experienced a flood during the past 5 years (Panel A) are estimated to have on average: a 2.9 percentage points higher likelihood of poor health (13.4% increase relative to the sample mean, $p = 0.228$); 0.132 more acute morbidities in the past four weeks (5.8% increase, $p = 0.098$); and a 0.634 higher depression score (8.1% increase, $p = 0.029$). Unsurprisingly, each of these effects are larger when the largest flood has occurred in the past year (Panel B). For such respondents, the poor health effect equals 7.6 percentage points (35% increase, $p = 0.068$), the acute morbidity effect equals 0.331 (15% increase, $p = 0.022$), and the depression effect equals 1.447 (19% increase, $p = 0.004$). The estimates for people who experienced a flood greater than one year ago are small and statistically insignificant, indicating that the negative health effects for adults dissipate over time.

In Columns (4) and (5) of Table 2, we present estimates for IFLS children. Similar to adults, children suffer more acute morbidities after floods: an average increase of 0.327 morbidities (15% increase, $p = 0.015$). A supplementary regression in Table A7 indicates that this effect is driven by increases in diarrhea, coughs and runny noses, which can lead to a decrease in anthropometric growth over the long term (Richard et al., 2013). Unlike adults, the estimated effects for parent-assessed 'poor health' are small, and the increase in morbidities appears similar for children who experienced a flood less than and greater than 1 year ago (0.322 versus 0.325). As noted previously, the IFLS does not include information on child mental health.

---

[12] Of the 15,902 households covered in IFLS5, 53.51% did not move between 2007 and 2014. Of the remaining 46.69% only 7.99% moved to another IFLS province or to a different province not surveyed by the IFLS. Despite this movement of households, only 9.31% of individuals surveyed in IFLS4 were not surveyed in IFLS5 of which 4.38% passed away and 4.93 left the sample. This highlights the substantial efforts of surveyors to track movers.

[Table 2]

Overall, the IFLS estimates indicate that floods worsen the physical health of adults and children, and worsen the mental health of adults. We next explore whether these estimated effects are driven by certain urban poor subpopulations. Table 3 presents estimated coefficients on 'flood in last 5 years' from regressions estimated separately by gender, age, and education. Panel (A) suggests that women and girls are more negatively affected by floods, with the estimated morbidity effects and the adult depression effect larger for females. However, none of the gender differences are statistically significant.

Regressions estimated separately by age group are presented in Panel B. The differences in acute morbidity effects are stark. The effects are large for older adults aged $\geq 41$ (0.351) and young children aged $\leq 7$ (0.490), and small for younger adults aged $\leq 40$ (-0.005) and older children aged 8-14 (0.023). The adult difference is statistically significant ($p = 0.017$), while the child difference is not ($p = 0.132$). The depression effect is also larger for older respondents (0.977 versus 0.668), but the difference is smaller and statistically insignificant. It's possible that these age differences are due to differences in time use. Younger children and older adults are likely to spend more time in or near their homes, and therefore have higher exposure to environmental contamination caused by flooding. Older children and younger adults are more likely to be at school or work and thereby avoid some portion of the negative flood effects. A similar mechanism may explain the observed gender patterns, given that women and girls spend more time in or near their homes than men and boys (UNICEF, 2020; Rubiano-Matulevich & Viollaz, 2019; OHCHR, 2017).

Finally, we turn to differences in flood effects by socio-economic status. In Panel C this is measured by educational attainment of the respondent for the adult sample, and educational attainment of the household head for the child sample. The results for adults suggest that the flood effects are larger for low SES households; but the evidence is weak. Lower educated adults experience worse physical health effects than do higher educated adults (0.109 versus - 0.005 for general health, and 0.248 versus 0.097 for acute morbidities). The results for mental health are somewhat different, where the estimated effects on depressive symptoms are not statistically significant for either group. For children, those from higher SES households experience more acute morbidities following floods than those from low SES households, yet the differences are statistically insignificant. The empirical evidence suggests that more educated mothers are likely to have healthier children due to better knowledge of health care and nutrition, and practice of healthier behaviors (Chen & Li, 2009). However, parents with higher education are also better at monitoring the health of their children and have higher health literacy. This could lead to better educated parents providing a more accurate account of child morbidities and less educated parents underreporting their children's health, thus explaining the findings from Panel C. Recall also that all of the households in our estimation sample are observed to be below the poverty line in either wave, and so these differences are identified by comparing people who are more and less severely poor.

[Table 3]

Thus far we have focussed on the urban poor because of their higher likelihood of experiencing negative health outcomes following disaster. In Appendix Table A8 we present estimates for the remaining surveyed households. Specifically, we separately estimate equation (1) for households in the higher third, fourth, and fifth quintiles of the household consumption distribution. The results suggest that the health of Indonesian adults in the top three quintiles is not significantly reduced in the years after a flood, with most point estimates in Table A8 smaller than their corresponding estimates in Table 2. Overall, these results suggest that households with greater economic means are typically better able to protect their health from the negative consequences of disaster events.

We also test the sensitivity of our results to the use of alternative 'urban poor' samples. In our main approach we have used consumption expenditure to define our sample of interest, but this variable is potentially affected by flood exposure. In Appendix Table A9 we explore how the estimates change when using alternative samples defined by: (i) residing in a village that is assessed by the village head in IFLS4 to have low financial prosperity; and (ii) residing in a village in which >30% of households received Raskin (a government rice subsidy) in the 12 months prior to the IFLS4 survey. Across the two alternative samples, the estimates are consistent with our main results. Generally, the coefficients decrease in magnitude, which is expected given the use of more imprecise (village-level) measures of household poverty, but remain statistically or economically significant in line with Table 2.

Another potential issue is that flood exposure is self-reported and therefore could suffer from reporting bias. We explore this possibility by testing whether certain respondents are more or less likely to report local flood events in their IFLS interview. Specifically, we regress the self-reported flood exposure indicator on a variable representing flood severity (number of villages flooded) in the past three years in the individual's district[13], interactions between this variable and individual-level characteristics (age, gender, education, general health, mental health), and individual and time fixed-effects. Economically and statistically significant interaction effects would suggest that the reporting of local floods differs substantially between people; which could be due to reporting bias. Results presented in Appendix Table A10 indicate that an individual's likelihood of reporting a flood in the IFLS is strongly related to the occurrence of floods within the local district, but importantly, this relationship is not moderated by individual-level characteristics. We fail to reject the null hypothesis that the interaction coefficients are jointly equal to zero ($p = 0.481$).

## 5. Health Effects of a Major Flood Event on Residents of Informal Settlements

The January 2019 flood in South Sulawesi was caused by monsoon rains and was one of the largest floods experienced by the province in recorded history, with some of the settlements recording up to 300 centimeters of rain in a 3-hour interval (Wolff, 2021). After the flood, 59 people were reported dead, 4,857 dwellings were flooded, and around 3,481 people were evacuated, leading to the Government declaring emergency status for the South Sulawesi province (IFRC, 2019). In this section we use data from the RISE household survey to estimate the impacts of this devastating event on physical and mental health outcomes of adults and children living in informal settlements.

In contrast to our analysis of the IFLS sample, our empirical specification for RISE residents uses reported flood damage (during the January 2019 flood), rather than flood exposure as the main treatment variable. A key advantage of this treatment variable in RISE compared to the IFLS is that it is less likely to suffer from recall bias, as individuals are asked to report damage in the days following the floods, instead of months and years. However, one potential concern is that the perception and therefore the reporting of damage from the January 2019 flood can be endogenous to the household's underlying vulnerability and ability to cope with the shock. If this were the case, the optimal strategy would be to exploit more exogenous flood data. However, due to the small geographical spread of the settlements within the city of Makassar and their close proximity to each other, this approach would essentially lead us to consider all RISE settlements as treated. Instead, we acknowledge the caveats of our treatment variable and, to control for the possibility that flood damage and its reporting was determined by household level factors – such as the household's insurance network, housing quality, the location of the house within a settlement – we estimate household fixed effects regressions:

$$hlth_{iht} = \delta_1(damage_h * w2_t) + \delta_2(damage_h * w3_t) + X'_{iht}\beta + \alpha_h + \varepsilon_{hst} \qquad (2)$$

where $hlth_{iht}$ is the health outcome of interest for individual $i$ living in household $h$ at time $t$, $damage_h$ is a binary variable indicating whether the flood event caused damage to the household's house, land or other assets, $w2_t$ and $w3_t$ indicate the two survey waves following the flood (waves 2 and 3), $\alpha_h$ is a household fixed effect, and $X_{iht}$ is a vector of covariates, including wave indicators and household characteristics, such as age of household head, and number of household members and children in the family. The coefficients of interest $\delta_1$ and $\delta_2$ measure the extent to which damage from the 2019 flood impacted health 4-5 months and 10-11 months after the flood, respectively.

---

[13] Flood data at the district level comes from the Village Potential Statistics (PODES) dataset for the period 1990 – 2019 (PODES, 2019). The variable we use is the percentage of villages within each district that experienced a flood in the 3 years prior to 2008 and 2014, the closest waves to our IFLS sample.

Before presenting estimated effects, we explore the randomness of flood damage across households within a settlement by regressing flood damage on pre-disaster characteristics.[14] Results are reported in Appendix Table A11. While there is some evidence suggesting that education may predict flood damage, the F-statistic of 0.83 for adults ($p = 0.643$) and 0.94 for children ($p = 0.5119$) indicates that pre-flood characteristics are unable to explain a significant proportion of the variation across households.

Panel A of Table 4 reports the estimated effects of the January 2019 flood on adult poor health, number of acute morbidities, and depression. The results show that 4-5 months after the flood event, direct flood damage was associated with a 12.7 percentage point increase in poor health (45.36% increase relative to the sample mean, $p = 0.064$), and 0.318 more acute morbidities (60% increase, $p = 0.022$). These effects do not persist over time, however, with the estimates for both physical health outcomes considerably smaller 10-11 months post-flood. For mental health, which was only measured in waves 1 and 3, we find that flood damage caused a 1.22-unit increase (18.8% relative to the mean, $p = 0.026$) in the adult depression score.

---

[14] Given that household fixed-effects are included in all regressions (i.e. a within-household identification approach), it isn't econometrically necessary for flooding to occur randomly across households. Nevertheless, it is informative to demonstrate the extent to which the major 2019 flooding event was indiscriminate in causing damage across houses as demonstrated in Appendix Table A12.

For the child sample in Panel B, we observe that flood damage increased poor health by 11.7 percentage points (68% increase, $p = 0.096$), and increased the number of acute morbidities by 0.22 (37.8%, $p = 0.108$); but both effects are imprecisely estimated. In terms of emotional functioning, children in households affected by flood damage experienced a large 13.57 unit (78% increase, $p = 0.000$) increase in their emotional problems score 4-5 months post flood. This is an indication of a substantial deterioration in mental health. This effect reduces in magnitude, but is still apparent and significant 10-11 months after the flood.

[Table 4]

To examine heterogeneity in effect sizes, we separately estimate equation 2 using similar sub-samples to those in the IFLS analysis (Table 5). For the adult sample, Panel B suggests that the health impacts are driven by people aged less than 40 years, where the difference in number of acute morbidities between both groups is statistically significant at the 10% level. This is in contrast to the IFLS findings where the older age group was most affected.

In terms of education, the results for poor health, and number of acute morbidities particularly, are driven by higher educated adults, but the difference between groups is not statistically significant for either variable ($p = 0.928$ and $p = 0.347$, respectively). The effect of flood damage on mental health is driven predominantly by the lower educated: with a 2.735 unit increase for lower education levels ($p = 0.001$) compared to 0.018 for higher education levels ($p = 0.978$). This difference is statistically significant ($p = 0.012$).

In columns 4 to 6 we examine whether similar patterns are present among the child sample. In terms of differential effects by gender (Panel A), we find a more consistent picture than that observed in the IFLS analysis; though not all differences are statistically significant. Girls are estimated to experience a six-times larger poor health effect ($p = 0.203$), a three-times larger acute morbidity effect ($p = 0.393$), and a two-times larger emotional functioning effect ($p = 0.077$). With regards to age (Panel B), the effects of flood damage on poor health are larger amongst younger children compared to older children (0.299 versus 0.037, $p = 0.029$). The impacts on acute morbidities and emotional problems, however, are similar across age groups.

Finally, we examine whether the education of a child's caregiver provides a protective effect from the shock. Similarly to the IFLS results, we find that the effect size for number of acute morbidities is significantly higher amongst children from more educated caregivers compared to children from less educated caregivers ($p = 0.002$). Rather than a lack of protective effects of parental education, we suggest that this reflects higher accuracy in reporting from caregivers with higher education. In contrast, effect sizes are similar for emotional problems and parent-assessed poor health.

[Table 5]

## 6. Economic Impacts of Floods

In this final empirical section, we explore the possibility that increased economic stress is a mechanism for the deleterious mental health effects documented in Tables 2 and 4. First, using IFLS data and the same covariate set as specified in regression equation (1), we estimate the effects of floods on various measures of household expenditure, and on three potential financial coping instruments: asset wealth, savings, and borrowing attempts. The estimates in Panel A of Table 6 show that medical expenditures[15] increased by a considerable 24.6% ($p = 0.069$), and non-medical expenditures increased by 6.5% ($p = 0.095$), for households that experienced flooding in the past year. The latter increase is primarily a result of increased food expenditure, which could reflect a positive price shock following the floods, rather than an increase in consumed calories.

In terms of how households cope with this expenditure shock, we first explore whether the flooding caused a reduction in the likelihood of having positive savings. The point estimates in Column 3 are relatively small but have large standard errors, and so provide little information about the role of savings. In contrast, the results in Column 4 suggest that flood victims experienced a reduction in assets: a 15.3% ($p = 0.022$) of a standard deviation decrease in the standardized wealth index, which was formed by using a principal component analysis on a list of both durable and non-durable assets. This reduction in asset wealth could have been directly caused by flood damage, but may also have been due to the forced selling off of assets. This latter explanation is supported by the results in Column 5. Flood exposure is estimated to significantly increase attempts to borrow (from a source other than family or friends) by 5.7 percentage points (33.5% increase with respect to baseline, $p = 0.084$), even for individuals

---

[15] Approximately 89% of households reported positive medical expenditures in the past month. Flood exposure did not significantly affect the likelihood of having a medical expenditure (-0.006, $p = 0.818$)

that experienced flooding more than one year prior. Altogether, these estimates suggest financial distress following floods may have compounded the direct health effects of flood exposure, and therefore partly explain the significant increase in adult depressive symptoms observed in Table 2.

[Table 6]

We next turn to RISE household respondents who report two measures of financial wellbeing: a) whether or not the household experienced a worsening in their finances in the past 6 months, and b) a continuous measure of the respondent's financial satisfaction on a scale of 0-10. In Panel B of Table 6, we employ these measures in a modified version of equation (2). Contrary to expectations, we find no significant effects of flood damage on subjective financial worsening ($p = 0.715$) or financial satisfaction ($p = 0.109$); though in the latter case the point estimates are relatively large. Based on these estimates, we are unable to conclude that there is a likely economic mechanism for the mental health effects presented in Table 4.

## 7. Discussion

In this study, we use the Indonesia Family Life Survey and a novel longitudinal study of informal settlement residents to investigate the effects of flood disaster exposure on the health of low-income adults and children living in urban areas of Indonesia. Specifically, we estimate the impact of flood victimization on the probability of being in poor health, the reporting of acute morbidities, and depression-related symptoms among adults. We also provide some of the first estimates of natural disaster effects on child emotional functioning in a low-income country setting. Further, the unique data also allow us to compare flood health effects across the urban poor living in formal housing, those living in informal settlements, and the rest of the consumption expenditure distribution including the urban wealthy.

Estimates from both urban poor samples indicate that flood disaster events increase the incidence of self-assessed poor health and acute morbidities among urban poor adults and children. The effects are strongest 4 to 5 months post flood, and dissipate by 10 to 11 months. Similarly, if the flood occurred 1 to 5 years prior to the survey, no detectable physical health effects remain. These findings highlight that frequent and proximal health sampling post-disaster may be important for obtaining reliable estimates of the health effects of floods.

Our findings also show that floods significantly increase adult depression-related symptoms and worsen child emotional functioning scores. The large deteriorations in child emotional functioning are consistent with other sources which indicate that children are at particularly high risk of severe mental impairment following natural disasters, because of their inability to understand the situation, acute feelings of helplessness, loss of attachments, and less experience coping with such type of situations (CDC, 2020; Norris et al., 2012). Particularly concerning is that, in contrast to the dissipating physical health effects, this mental health toll persists in both adults and children at 10 to 11 months in informal settlements, as well as on the sample of general urban poor who experienced the flood 1 to 5 years prior. The evidence also suggests that the mental health impacts among the IFLS sample could be driven by flood-induced financial distress.

Our study also provides insights into the wealth-gradient in flood health effects. Our estimates indicate that the physical health and mental health of medium to high socioeconomic status households (the top three quintiles of the urban consumption distribution) were mostly unaffected by flood events, whereas those living in informal settlements and in the bottom two quintiles experience substantial deteriorations across all health outcomes. A possible explanation for this is raised by Ezeh et al. (2017) and Lilford et al. (2017), who describe slums as unique settings due to the shared physical and social environment. The inadequate water supply, garbage collection, sewage disposal and sanitation resulting from being out of the grid, and living in tightly packed clusters, might serve to predispose residents to infectious diseases and then exacerbate the spread of such diseases when a flood occurs. Such findings emphasize the importance of focusing on the urban poor both as a subject of study as well as a key population on which to focus policy action. They also provide support for policy that moves towards the formal recognition of informal settlements and the expansion of public services in poor communities to curb the spread of diseases.

Currently one in four Indonesians live in high-risk flood zones due to poor planning in fast urbanizing areas (Rentschler et al., 2021). The expansion of housing settlements along the catchment areas of rivers and on deforested land, along with poor management of waste and water infrastructure, contribute to an increased susceptibility to flooding across Indonesian cities (JBA Risk Management, 2021). For example, in Jakarta, the capital of Indonesia and the largest city in Southeast Asia, most of its residents rely on wells to obtain fresh water supply which is contributing to the sinking of the city (Business Insider, 2019). As a result, the city is not only vulnerable to flood risk from above average rainfall but also to climate change related sea level rise (WSJ, 2020). A water sensitive approach to urban planning and design that mimics the natural water cycle and anticipates the way in which urbanization disrupts water flow could help reduce this risk (Melbourne Water, 2017).

The results of this study are also a call for action. There is already a need to enhance the preparedness and capacity of health and emergency response systems to cope with the challenges faced by flood affected adults and children (Power et al., 2017). As climate change worsens, more and more people globally will face major floods for the first time, and this need will only increase. Our estimates show that the poor in urban areas of developing countries, and paradoxically those who contribute least to climate change, will likely experience the largest toll. Nevertheless, the lack of negative health effects on the middle and top quintiles of the consumption expenditure distribution suggests that this course is not irreversible and that the poor can be spared too if their communities can be supported to become more resilient and post-disaster recovery targeting is improved.

# Figures

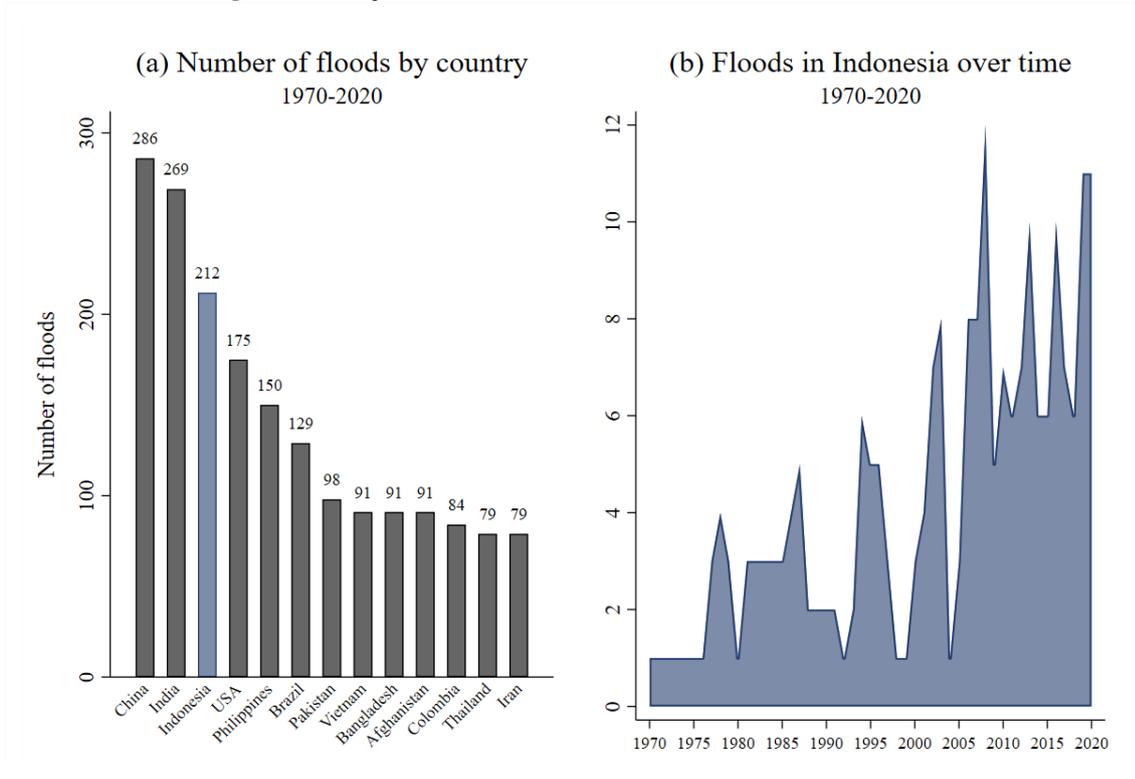

**Figure 1.** Major Flood Occurrences in Indonesia and the World

**Note.** - Graphs created by authors based on flood occurrence data by country available at the Emergency Events Database (EM-DAT) by the Université Catholique de Louvain (UCL), CRED, D. Guha-Sapir. Disasters qualify into the database if they fulfill at least one of the following criteria: 10 or more people reported killed, 100 or more people reported affected, a declaration of a state of emergency or a call for international assistance (https://www.emdat.be/explanatory-notes). Figure 1a includes the top 13 countries in the world in terms of flood occurrence in the past 50 years. Figure 1b shows the number of large-scale floods reported every year since 1970 by the Indonesian authorities.

## Tables

**Table 1.** Key Variables used in the IFLS and RISE Regression Analyses

| Variables | Description | Adult sample mean | Child sample mean |
|---|---|---|---|
| (A) IFLS | | | |
| Flood | Household experienced a flood in the last 5 years | 10.35% | 11.85% |
| Poor health | General health is somewhat unhealthy or very unhealthy | 21.60% | 11.22% |
| Acute morbidities | Number of morbidities in past 4 weeks: runny nose, cough, stomach ache, nausea, diarrhea, skin infections | 2.26 | 2.22 |
| Depression score | CESD-10 score of the frequency of 10 depression-related feelings during the past week. E.g. "I felt depressed", "I felt fearful". Ranges from 0-30, where higher scores indicate worsening of depressive symptoms and a score of 10+ indicates a person may have depression. | 7.80 | - |
| Log Non-Medical Expenditures | Log of all monthly expenditures made by the household, except for medical expenditures (e.g. food, rent, educ.) | 12.90 | |
| Savings | An indicator that the household member owns savings | 22% | - |
| Asset wealth | A standardized score of asset wealth constructed using PCA of a combination of durable non-durable assets | 0 | - |
| Try to borrow | An indicator that member tried to borrow money or goods from a source other than family or friends in the past 12 months | 17% | - |
| (B) RISE | | | |
| Flood damage | January 2019 flood caused property damage | 28.83% | 28.85% |
| Poor health | General health is very bad, bad or moderate | 28% | 17% |
| Acute morbidities | Number of morbidities in the past 1 week: cough, difficulty breathing, diarrhea, fever, skin infections | 0.53 | 0.59 |
| Depression score | CESD-10 score representing frequency of 10 depression-related feelings during the past week. E.g. "I felt depressed", "I felt fearful". Ranges from 0-30, where higher scores indicate worsening of depressive symptoms and a score of 10 or higher indicates a person may have depression. | 6.49 | - |
| Emotional problems score | PedsQL Emotional Functioning score representing frequency of 5 feelings during the past week. E.g. "feeling sad or blue", "feeling afraid or scared". Ranges from 0-100, where higher scores indicate worse emotional functioning. | - | 17.38 |
| Worse finances | An indicator that the household experienced a worsening in their financial situation in the 6 months prior to survey | 7% | - |
| Financial satisfaction | A continuous score between 0-10 indicating the respondent's level of financial satisfaction | 5.95 | - |

**Note.** - IFLS sample means for adults and children were computed using the combined samples from IFLS4 and IFLS5. RISE sample means for adults and children were computed using wave one data prior to the flood.

**Table 2.** Estimated effects of floods for poor IFLS respondents living in urban areas

|  | Adults | | | Children | |
| --- | --- | --- | --- | --- | --- |
| Variables | Poor Health | Acute Morbidities | Depression Score | Poor Health | Acute Morbidities |
|  | (1) | (2) | (3) | (4) | (5) |
| (A) Flood in last 5 years | 0.029 | 0.132* | 0.634** | 0.012 | 0.327** |
|  | (0.024) | (0.079) | (0.290) | (0.031) | (0.134) |
| (B) Flood 0-1 year ago | 0.076* | 0.331** | 1.447*** | 0.013 | 0.322 |
|  | (0.042) | (0.144) | (0.496) | (0.059) | (0.251) |
| Flood > 1 year ago | 0.019 | 0.0921 | 0.466 | 0.011 | 0.325** |
|  | (0.027) | (0.086) | (0.317) | (0.033) | (0.144) |
| Mean Outcome | 21.60% | 2.26 | 7.80 | 11.22% | 2.22 |
| Observations | 8006 | 8006 | 7591 | 4072 | 4072 |

**Note.** - All regressions control for individual fixed effects, wave dummies, individual and household characteristics, and province*wave fixed effects. Individual characteristics include a cubic function of age. Household characteristics include age of household head, age of head squared, number of children under 14 in the household, and number of members per household. Individuals were considered poor if their average daily equivalized income in IFLS4 and IFLS5 was below the $1.51 ADB poverty line for Asia Pacific countries. All models use data from waves 4 and 5 of the IFLS. Robust standard errors in (parentheses). * $p < 0.1$; ** $p < 0.05$; *** $p < 0.01$.

**Table 3**. Estimated effects of floods for subsamples of IFLS respondents

| Variables | Adults | | | Children | |
|---|---|---|---|---|---|
| | Poor Health | Acute Morbidities | Depression Score | Poor Health | Acute Morbidities |
| | (1) | (2) | (3) | (4) | (5) |
| (A) Gender | | | | | |
| Males | 0.055 | 0.015 | 0.383 | 0.076** | 0.309 |
| | (0.036) | (0.123) | (0.474) | (0.039) | (0.193) |
| Females | 0.011 | 0.237** | 0.878** | -0.045 | 0.378** |
| | (0.032) | (0.106) | (0.354) | (0.047) | (0.184) |
| (B) Age | | | | | |
| Younger | -0.010 | -0.005 | 0.668 | 0.005 | 0.490*** |
| | (0.033) | (0.126) | (0.452) | (0.044) | (0.184) |
| Older | 0.058 | 0.351*** | 0.977** | 0.009 | 0.023 |
| | (0.042) | (0.112) | (0.447) | (0.040) | (0.203) |
| (C) Education | | | | | |
| Primary or lower | 0.109*** | 0.248** | 0.796 | 0.001 | 0.054 |
| | (0.039) | (0.123) | (0.511) | (0.051) | (0.213) |
| At least high school | -0.005 | 0.097 | 0.524 | 0.013 | 0.538*** |
| | (0.033) | (0.117) | (0.365) | (0.051) | (0.222) |
| Observations | 8006 | 8006 | 7591 | 4072 | 4072 |

**Note.** – All subsample analyses are performed using the binary indicator of flood at any point in the five years prior to the survey. All regressions control for individual FE, wave dummies, individual and household characteristics, as well as province*wave FE. Individual characteristics include a cubic function of age. Household characteristics include age of household head, age of head squared, number of children < 14 in the household, and number of members per household. Young adults in panel B are age 40 years or less while older adults are 40+. Young children are in the 0-7 years age group, while older children are in the 8-15 age group. For panel C, columns 1 to 3 use the education of the household member and columns 4 and 5 use the education of the household head to divide the children sample. Those with higher education have at least one year of high-school. All models use data from waves 4 and 5 of the IFLS. Robust errors in (parentheses).

* p < 0.1; ** p < 0.05; *** p < 0.01.

**Table 4.** Estimated effects of damage from the January 2019 flood for RISE respondents in informal settlements of Makassar, Indonesia

| Variables | Poor Health (1) | Acute Morbidities (2) | Depression Score / Emotional Problems (3) |
|---|---|---|---|
| (A) Adults | | | |
| 4-5 months after flood | 0.127* | 0.318** | - |
|  | (0.068) | (0.138) |  |
| 10-11 months after flood | 0.026 | 0.084 | 1.221** |
|  | (0.070) | (0.139) | (0.545) |
| Mean Outcome | 28% | 0.53 | 6.49 |
| Observations | 1429 | 1429 | 897 |
| (B) Children | | | |
| 4-5 months after flood | 0.117* | 0.223 | 13.575*** |
|  | (0.070) | (0.138) | (3.094) |
| 10-11 months after flood | -0.006 | -0.062 | 5.232* |
|  | (0.056) | (0.141) | (2.673) |
| Mean Outcome | 17% | 0.59 | 17.38 |
| Observations | 1716 | 1715 | 1030 |

**Note.** – Regression results are based on equation 2 using household fixed effects. All models control for individual and household characteristics as well as wave and household fixed effects, and settlement*wave interactions. Individual characteristics include: a cubic function of age and gender. Household characteristics include: number of children and number of people in the household. House characteristics include: a list of all assets in the house at baseline, material of floor, roof and walls. All models include data from three waves: Baseline, Wave 2 (4-5 months post flood) and Wave 3 (10-11 months post flood). Robust standard errors in (parentheses).
* $p<0.1$; ** $p<0.05$; *** $p<0.01$.

**Table 5.** Estimated effects of January 2019 flood for subsamples of RISE respondents

| Variables | Adults | | | Children | | |
|---|---|---|---|---|---|---|
| | Poor Health (1) | Acute Morbidities (2) | Depression Score (3) | Poor Health (4) | Acute Morbidities (5) | Emotional Problems (6) |
| (A) Gender | | | | | | |
| Males | - | - | - | 0.029 | 0.102 | 7.863* |
| | | | | (0.087) | (0.204) | (4.397) |
| Females | 0.143* | 0.291** | 0.952 | 0.174* | 0.321* | 17.807*** |
| | (0.076) | (0.146) | (0.586) | (0.093) | (0.172) | (4.311) |
| (B) Age | | | | | | |
| Younger | 0.108 | 0.538*** | 1.394** | 0.299*** | 0.198 | 13.243*** |
| | (0.092) | (0.158) | (0.668) | (0.099) | (0.219) | (3.974) |
| Older | 0.024 | 0.042 | 0.135 | 0.037 | 0.223 | 13.059** |
| | (0.144) | (0.356) | (1.243) | (0.085) | (0.180) | (5.433) |
| (C) Education | | | | | | |
| Primary or lower | 0.083 | 0.258 | 2.735*** | 0.137 | -0.326 | 15.512*** |
| | (0.125) | (0.262) | (0.837) | (0.111) | (0.231) | (5.833) |
| At least high school | 0.097 | 0.561*** | 0.018 | 0.097 | 0.584*** | 15.114*** |
| | (0.103) | (0.187) | (0.669) | (0.094) | (0.169) | (4.164) |
| Observations | 1429 | 1429 | 897 | 1716 | 1715 | 1030 |

Note. - Regression results are based on equation 2 using household fixed effects. All regressions control for individual, household and house characteristics as well as wave and household fixed effects, and settlement*wave interactions. Individual characteristics include age and gender except for panel A where age is rather a dummy interacted with damage. Household characteristics include the number of children and people in the household. House characteristics include controls for all assets in the house at baseline, material of the floor, roof and walls at baseline. Young adults in panel B are age 40 years or less while older adults are 40+, while young children are 0-7 years old and older ones are 8-15. In panel C, those with higher education have at least one year of high-school. Robust standard errors in (parentheses).
* p<0.1; ** p<0.05; *** p<0.01.

**Table 6.** Estimated effects of flood on financial wellbeing and expenditures in IFLS and RISE

| Variables | IFLS | | | | | RISE | |
|---|---|---|---|---|---|---|---|
| | Log Non-Med Expend | Log Medical Expend | Savings | Asset Wealth | Try to Borrow | Worse Finances | Financial Satisfaction |
| | (1) | (2) | (3) | (4) | (5) | (6) | (7) |
| (A) Flood 0-1 year ago | 0.065* | 0.246* | -0.016 | -0.153** | 0.057* | | |
| | (0.039) | (0.136) | (0.032) | (0.067) | (0.033) | | |
| Flood > 1 year ago | -0.023 | -0.058 | 0.018 | -0.010 | 0.057*** | | |
| | (0.019) | (0.067) | (0.016) | (0.034) | (0.017) | | |
| (B) 4-5 months after | | | | | | -0.010 | 0.486 |
| | | | | | | (0.028) | (0.303) |
| 10-11 months after | | | | | | -0.045 | 0.228 |
| | | | | | | (0.035) | (0.290) |
| Mean Outcome | 12.90 | 9.41 | 0.22 | 0 | 0.17 | 0.07 | 5.95 |
| Observations | 9805 | 8724 | 9805 | 9804 | 9805 | 1429 | 1406 |

Note. – Regression results in panel A are based on equation 1 using the IFLS data and results from panel B are based on equation 2 using RISE data. Outcomes in columns 1 and 2 are in logarithmic values. Robust standard errors in (parentheses). * p<0.1; ** p<0.05; *** p<0.01.

# Appendix

Table A1. Descriptive statistics of IFLS respondents that are poor in urban areas of Indonesia

| Variables | Adult Sample | | Child Sample | |
|---|---|---|---|---|
| | IFLS4 | IFLS5 | IFLS4 | IFLS5 |
| | (1) | (2) | (3) | (4) |
| **Demographic Characteristics** | | | | |
| Age now | 39.66 | 46.50 | 6.66 | 13.76 |
| Gender: Male | 0.46 | 0.45 | 0.51 | 0.49 |
| Marital Status - Unmarried / Head (children) | 0.24 | 0.12 | - | - |
| Marital Status - Married / Head (children) | 0.62 | 0.69 | 0.84 | 0.77 |
| Marital Status - Separated / Head (children) | 0.01 | 0.01 | 0.03 | 0.03 |
| Marital Status - Divorced / Head (children) | 0.03 | 0.04 | 0.12 | 0.11 |
| Marital Status - Widow / Head (children) | 0.10 | 0.14 | 0.02 | 0.08 |
| Household member has 6th grade or < | 0.48 | 0.48 | 0.81 | 0.40 |
| Still in School | - | - | 0.50 | 0.73 |
| Main activity: Working | 0.56 | 0.58 | 0.01 | 0.15 |
| Main activity: Looking for work | 0.01 | 0.01 | 0.00 | 0.01 |
| Main activity: Student | 0.06 | 0.00 | 0.81 | 0.71 |
| Main activity: Housekeeper | 0.23 | 0.27 | 0.00 | 0.05 |
| Main activity: Retired | 0.05 | 0.05 | 0.00 | 0.00 |
| Main activity: Unemployed | 0.08 | 0.06 | 0.18 | 0.07 |
| Main activity: Sick | 0.01 | 0.03 | 0.00 | 0.01 |
| Relation to head: Children (biological) | 0.24 | 0.15 | 0.68 | 0.67 |
| Relation to head: Children (step/adopted) | 0.01 | 0.01 | 0.02 | 0.03 |
| Relation to head: Grandchild | 0.02 | 0.01 | 0.27 | 0.13 |
| Relation to head: Nephews/nieces | 0.01 | 0.00 | 0.02 | 0.03 |
| Relation to head: Other | 0.72 | 0.84 | 0.01 | 0.13 |
| Age of household head | 50.96 | 51.39 | 45.32 | 45.31 |
| Household head has 6th grade or < | 0.45 | 0.38 | 0.39 | 0.39 |
| Activity of head: Working | 0.72 | 0.71 | 0.78 | 0.75 |
| Activity of head: Housekeeper | 0.09 | 0.11 | 0.09 | 0.10 |
| Activity of head: Retired | 0.09 | 0.07 | 0.04 | 0.03 |
| Activity of head: Unemployed | 0.08 | 0.07 | 0.06 | 0.06 |
| Activity of head: Sick | 0.02 | 0.03 | 0.01 | 0.02 |
| Expenditure quintile: Poorest | 0.44 | 0.53 | 0.43 | 0.48 |
| Expenditure quintile: Poorer | 0.41 | 0.22 | 0.44 | 0.23 |
| Expenditure quintile: Middle | 0.08 | 0.12 | 0.06 | 0.13 |
| Expenditure quintile: Richer | 0.05 | 0.08 | 0.05 | 0.09 |
| Expenditure quintile: Richest | 0.02 | 0.05 | 0.02 | 0.06 |
| Material used in walls is non-porous | 0.79 | 0.85 | 0.77 | 0.84 |
| Floor is made of dirt or unfinished | 0.05 | 0.03 | 0.05 | 0.03 |
| Number of children < 14 per household | 1.28 | 1.17 | 2.32 | 1.59 |
| Number of members per household | 4.65 | 4.34 | 5.58 | 4.95 |
| **Outcome Variables** | | | | |
| Self-reported unhealthy | 0.16 | 0.28 | 0.11 | 0.11 |
| Reported at least 1 Morbidity in past 4 weeks | 0.72 | 0.81 | 0.67 | 0.80 |
| Number of Morbidities in past 4 weeks | 1.93 | 2.51 | 1.98 | 2.40 |
| Cough in the past 4 weeks | 0.32 | 0.42 | 0.38 | 0.43 |
| Difficulty breathing in the past 4 weeks | 0.07 | 0.09 | 0.03 | 0.05 |
| Stomach ache in the past 4 weeks | 0.19 | 0.29 | 0.17 | 0.34 |
| Nausea in the past 4 weeks | 0.10 | 0.16 | 0.10 | 0.13 |
| Diarrhea in the past 4 weeks | 0.07 | 0.11 | 0.11 | 0.09 |
| Skin infection in the past 4 weeks | 0.09 | 0.15 | 0.07 | 0.12 |
| CES-D 10 Scale | 6.26 | 8.51 | - | - |
| Individual is depressed | 0.13 | 0.36 | - | - |
| Observations | 4878 | 3950 | 1997 | 2379 |

**Note.** - Mean statistics reported

**Table A2.** Descriptive statistics for RISE adults and children at baseline in informal settlements of Makassar, Indonesia

| Variables | Adult Sample (1) | Child Sample (2) |
|---|---|---|
| **Demographic Characteristics** | | |
| Age | 41.47 | 6.51 |
| Gender: Male | 0.10 | 0.53 |
| Number of people in household | 4.78 | 5.57 |
| Children under 15 | 1.42 | 2.21 |
| Lived <6 months in settlement | 0.01 | 0.01 |
| Lived 6 months-1 year | 0.02 | 0.01 |
| Lived 1-2 years in settlement | 0.04 | 0.04 |
| Lived 2-5 years in settlement | 0.09 | 0.11 |
| Lived 5-10 years in settlement | 0.12 | 0.16 |
| Lived >10 years in settlement | 0.44 | 0.41 |
| Lived all life in settlement | 0.28 | 0.25 |
| Freehold Ownership Title | 0.50 | 0.49 |
| Sale & Purchase Deed | 0.27 | 0.27 |
| Temporary Registration Letter | 0.11 | 0.12 |
| Right to Work Land | 0.01 | 0.02 |
| Proof of Payment/Instalment | 0.03 | 0.04 |
| Relocation Letter | 0.02 | 0.01 |
| Tenure - Other | 0.05 | 0.04 |
| Roofing - Corrugated tin / iron / alum / zinc | 1.00 | 1.00 |
| Material used in walls is porous | 0.53 | 0.48 |
| Walls - Masonry | 0.72 | 0.75 |
| Walls - Wood or plywood | 0.30 | 0.25 |
| Walls - Bamboo woven or mat | 0.02 | 0.02 |
| Walls - Tin or corrugated iron | 0.39 | 0.35 |
| Flood is made of dirt or unfinished | 0.42 | 0.37 |
| Flooring - Ceramic / tiles / terrazzo | 0.57 | 0.59 |
| Flooring - Laminate (plastic) | 0.10 | 0.09 |
| Flooring - Concrete | 0.46 | 0.48 |
| Flooring - Wood/boards | 0.41 | 0.36 |
| Flooring - Bamboo | 0.01 | 0.01 |
| Flooring - Soil/dirt | 0.03 | 0.03 |
| Open Sewage | 0.10 | 0.09 |
| Unprotected Water | 0.54 | 0.56 |
| In past 3 months flood outside/under house | 0.11 | 0.12 |
| **Outcome Variables** | | |
| Poor health of respondent/child | 0.39 | 0.22 |
| In last month, how many days where you sick to do normal act. | 1.53 | - |
| Respondent had at least 1 acute morbidity | 0.42 | 0.44 |
| Respondent number of morbidities | 0.64 | 0.69 |
| In last 3 months, have you seen health worker | 0.25 | 0.33 |
| In the last week, have you had a cough lasting through day? | 0.16 | 0.19 |
| In the last week, have you had trouble breathing? | 0.07 | 0.02 |
| In the last week, have you had a fever? | 0.11 | 0.21 |
| In the last week, did you have three or + loose stools in 24h? | 0.06 | 0.08 |
| In the last week, how many days did you have 3+ loose stools? | 0.14 | 0.20 |
| In the last week, have you had a skin infection? | 0.12 | 0.09 |
| CESD Depression 10 Scale / PedsQL Score | 6.70 | 18.74 |
| Individual is depressed | 0.18 | - |
| Observations | 500 | 579 |

**Note. -** Mean statistics reported at baseline

**Table A3.** Predicting the probability of attrition in IFLS5

| Variables | Attrition | Attrition: Passed Away | Attrition: Left sample |
|---|---|---|---|
| | (1) | (2) | (3) |
| Flood in IFLS4 | -0.003 | -0.002 | -0.001 |
| | (0.006) | (0.004) | (0.005) |
| Gender: Male | 0.022*** | 0.024*** | -0.002 |
| | (0.003) | (0.002) | (0.003) |
| Age of respondent in 2007 | 0.006*** | 0.005*** | 0.001*** |
| | (0.000) | (0.000) | (0.000) |
| Working in 2007 | -0.073*** | -0.058*** | -0.015*** |
| | (0.004) | (0.003) | (0.003) |
| High Education | -0.006** | -0.012*** | 0.006*** |
| | (0.003) | (0.002) | (0.002) |
| Relation to head: Spouse | -0.013** | -0.008** | -0.005 |
| | (0.005) | (0.004) | (0.004) |
| Relation to head: Children | 0.058*** | 0.047*** | 0.011** |
| | (0.006) | (0.004) | (0.005) |
| Relation to head: Other | 0.076*** | 0.071*** | 0.005 |
| | (0.007) | (0.006) | (0.005) |
| Married in 2007 | -0.028*** | -0.017*** | -0.011*** |
| | (0.005) | (0.004) | (0.004) |
| Log Consumption Expenditure | 0.025*** | -0.005*** | 0.030*** |
| | (0.003) | (0.002) | (0.002) |
| Number of household Members | -0.009*** | -0.002** | -0.007*** |
| | (0.001) | (0.001) | (0.001) |
| Number of children in household | 0.010*** | 0.003** | 0.007*** |
| | (0.002) | (0.001) | (0.001) |
| Province FE | YES | YES | YES |
| Mean Attrition | 9.31% | 4.38% | 4.93% |
| R-squared | 0.129 | 0.166 | 0.038 |
| Observations | 38831 | 38831 | 38831 |

**Note. -** Model 1 controls for all covariates included in the main regression results plus province fixed effects and indicators of socio-economic status such as working status, level of education and log of consumption expenditure. Models 2 and 3 perform a similar analysis but disaggregating the type of attrition by death and leaving the sample or unable to track. Reference levels are relation to head (Head of Household), Marital Status (Single, Divorced, Widowed) High Education (Less than 1 year of secondary education). Robust standard errors in (parentheses).
* $p<0.1$; ** $p<0.05$; *** $p<0.01$

**Table A4.** IFLS Results for IFLS urban poor adults and children in Indonesia with added (potentially endogenous) controls

| Variables | Adults | | | Children | |
|---|---|---|---|---|---|
| | Poor Health | Number of Morbidities | CESD Score | Poor Health | Number of Morbidities |
| | (1) | (2) | (3) | (4) | (5) |
| (A) Flood in past 5 years | 0.034 | 0.146* | 0.641** | 0.015 | 0.382*** |
| | (0.024) | (0.080) | (0.290) | (0.032) | (0.140) |
| (B) Flood 0-1 Year Ago | 0.081* | 0.329** | 1.475*** | 0.006 | 0.410 |
| | (0.042) | (0.147) | (0.502) | (0.060) | (0.255) |
| Flood > 1 Year Ago | 0.024 | 0.108 | 0.467 | 0.017 | 0.373** |
| | (0.027) | (0.086) | (0.318) | (0.035) | (0.150) |
| Mean Outcome | 21.60% | 2.26 | 7.8 | 11.22% | 2.22 |
| Observations | 8004 | 8002 | 7590 | 4048 | 4045 |

**Note.** - For the adult sample in columns 1-3, the additional controls include marital status, equivalized consumption, education level, and main activity of the household member. For the child sample in columns 4-5 we instead control for marital status, education level and primary activity of the child's caregiver and household wealth index. Robust standard errors in (parentheses).
* p<0.1; ** p<0.05; *** p<0.01

**Table A5.** Estimated effects of floods for urban poor IFLS (Regency Fixed Effects)

| Variables | Adults | | | Children | |
|---|---|---|---|---|---|
| | Poor Health | Acute Morbidities | Depression Score | Poor Health | Acute Morbidities |
| | (1) | (2) | (3) | (4) | (5) |
| (A) Flood in last 5 years | 0.023 | 0.083 | 0.490* | 0.018 | 0.282** |
| | (0.024) | (0.081) | (0.296) | (0.031) | (0.134) |
| (B) Flood 0-1 year ago | 0.058 | 0.318** | 1.312** | 0.022 | 0.289 |
| | (0.039) | (0.147) | (0.517) | (0.058) | (0.256) |
| Flood > 1 year ago | 0.015 | 0.037 | 0.325 | 0.017 | 0.279* |
| | (0.027) | (0.087) | (0.323) | (0.033) | (0.143) |
| Mean Outcome | 21.60% | 2.26 | 7.8 | 11.22% | 2.22 |
| Observations | 8006 | 8004 | 7591 | 4072 | 4072 |

**Note.** - All regressions control for individual fixed effects, wave dummies, individual and household characteristics, and regency fixed effects. Individual characteristics include a cubic function of age. Household characteristics include age of household head, age of head squared, number of children under 14 in the household, number of members per household. Individuals were considered poor if their average daily equivalized income in IFLS4 and IFLS5 was below the $1.51 ADB poverty line for Asia Pacific countries. All models use data from waves 4 and 5 of the IFLS. Robust standard errors in (parentheses).
* p < 0.1; ** p < 0.05; *** p < 0.01.

**Table A6.** Checking for Endogeneity of Floods among the IFLS Adult Sample

| Variables | Flood (1) |
|---|---|
| Log Consumption Expenditure | -0.001 |
|  | (0.009) |
| Poor Health | 0.009 |
|  | (0.015) |
| Number of Morbidities | 0.000 |
|  | (0.004) |
| Marital Status: Married | 0.012 |
|  | (0.028) |
| Marital Status: Separated | -0.019 |
|  | (0.064) |
| Marital Status: Divorced | -0.025 |
|  | (0.050) |
| Marital Status: Widow | 0.005 |
|  | (0.044) |
| Activity: Looking for work | 0.027 |
|  | (0.047) |
| Activity: Student | -0.014 |
|  | (0.023) |
| Activity: Housekeeper | 0.028* |
|  | (0.016) |
| Activity: Retired | 0.002 |
|  | (0.023) |
| Activity: Unemployment | -0.032 |
|  | (0.021) |
| Activity: Sick | -0.021 |
|  | (0.051) |
| Activity: Other | -0.051 |
|  | (0.067) |
| High Education | 0.008 |
|  | (0.020) |
| Age | 0.001 |
|  | (0.003) |
| Number of household Members | -0.001 |
|  | (0.003) |
| Number of children in household | 0.000 |
|  | (0.007) |
| Province-Wave FE | YES |
| Observations | 7064 |
| R-squared | 0.042 |
| F-Test | 0.61 |
| p-value | 0.893 |

**Note**. – To check for endogeneity of floods among the IFLS sample we regress floods in IFLS5 on health status, individual and household characteristics in IFLS4. Reported F-statistic tests for joint significance include all lags in table A6, except province-wave interactions. Robust standard errors in (parentheses).
* $p<0.1$; ** $p<0.05$; *** $p<0.01$.

**Table A7.** Estimated effects of floods on morbidities for poor IFLS respondents in urban areas of Indonesia

| Variables | Runny Nose (1) | Cough (2) | Stomach Ache (3) | Nausea (4) | Diarrhea (5) | Skin Infections (6) |
|---|---|---|---|---|---|---|
| (A) Adults | | | | | | |
| Flood in the last 5 years | 0.048 | 0.028 | 0.026 | 0.015 | 0.016 | 0.000 |
| | (0.031) | (0.028) | (0.027) | (0.021) | (0.018) | (0.020) |
| Flood 0-1 year ago | 0.058 | 0.078 | 0.074 | 0.078** | 0.017 | 0.026 |
| | (0.064) | (0.056) | (0.053) | (0.036) | (0.037) | (0.043) |
| Flood > 1 year ago | 0.045 | 0.018 | 0.016 | 0.002 | 0.015 | -0.005 |
| | (0.033) | (0.029) | (0.029) | (0.022) | (0.019) | (0.022) |
| Observations | 8004 | 8004 | 8004 | 8004 | 8004 | 8004 |
| (B) Children | | | | | | |
| Flood in the last 5 years | 0.079* | 0.120** | 0.025 | 0.000 | 0.067*** | 0.035 |
| | (0.047) | (0.049) | (0.041) | (0.030) | (0.026) | (0.028) |
| Flood 0-1 year ago | 0.073 | 0.205*** | -0.034 | 0.000 | 0.073 | 0.005 |
| | (0.082) | (0.076) | (0.082) | (0.056) | (0.054) | (0.060) |
| Flood > 1 year ago | 0.079 | 0.097* | 0.041 | 0.000 | 0.065** | 0.043 |
| | (0.051) | (0.054) | (0.042) | (0.032) | (0.029) | (0.031) |
| Observations | 4070 | 4070 | 4070 | 4070 | 4070 | 4070 |

**Note.** - Individual characteristics controlled for include a cubic function of age. Household characteristics controlled for include age of household head, age of head squared, number of children under 14 in the household, number of members per household. All morbidities reportedly occurred in the 4 weeks prior to the interview. All models use data from waves 4 and 5 of the IFLS. Robust errors in (parentheses).
* $p < 0.1$; ** $p < 0.05$; *** $p < 0.01$.

**Table A8.** Estimated effects of floods by quintiles of IFLS urban consumption expenditure
(Top 3 quintiles)

|  | Adults | | | Children | |
|---|---|---|---|---|---|
| Variables | Poor Health | Acute Morbidities | Depression Score | Poor Health | Acute Morbidities |
|  | (1) | (2) | (3) | (4) | (5) |
| (A) Quintile 3 | | | | | |
| Flood in last 5 years | 0.048 | -0.156 | 0.455 | 0.046 | -0.022 |
|  | (0.032) | (0.099) | (0.339) | (0.034) | (0.145) |
| Flood 0-1 year ago | 0.039 | -0.281 | 0.353 | 0.008 | 0.233 |
|  | (0.057) | (0.193) | (0.594) | (0.077) | (0.252) |
| Flood > 1 year ago | 0.050 | -0.127 | 0.480 | 0.053 | -0.078 |
|  | (0.036) | (0.108) | (0.379) | (0.035) | (0.158) |
| (B) Quintile 4 | | | | | |
| Flood in last 5 years | 0.012 | 0.092 | 0.079 | 0.048 | 0.148 |
|  | (0.025) | (0.082) | (0.270) | (0.031) | (0.116) |
| Flood 0-1 year ago | 0.026 | 0.368** | 0.635 | -0.079 | 0.365 |
|  | (0.042) | (0.171) | (0.502) | (0.064) | (0.295) |
| Flood > 1 year ago | 0.010 | 0.050 | -0.013 | 0.069** | 0.113 |
|  | (0.027) | (0.085) | (0.289) | (0.034) | (0.127) |
| (C) Quintile 5 | | | | | |
| Flood in last 5 years | 0.015 | 0.021 | 0.339 | 0.027 | 0.034 |
|  | (0.026) | (0.083) | (0.266) | (0.034) | (0.118) |
| Flood 0-1 year ago | 0.030 | 0.310* | 0.781 | 0.039 | -0.052 |
|  | (0.053) | (0.175) | (0.519) | (0.066) | (0.307) |
| Flood > 1 year ago | 0.012 | -0.027 | 0.264 | 0.025 | 0.047 |
|  | (0.027) | (0.086) | (0.282) | (0.035) | (0.125) |

**Note.** - All regressions control for individual fixed effects, wave dummies, individual and household characteristics, and province*wave fixed effects. Individual characteristics include a cubic function of age. Household characteristics include age of household head, age of head squared, number of children under 14 in the household, number of members per household. All models use data from waves 4 and 5 of the IFLS. Robust standard errors in (parentheses).
* $p < 0.1$; ** $p < 0.05$; *** $p < 0.01$.

**Table A9.** Estimated effects of floods on health on the IFLS urban sample using different poverty definitions

| Variables | Adults | | | | Children | |
|---|---|---|---|---|---|---|
| | Acute Morbidities | | Depression Score | | Acute Morbidities | |
| | Low Prosperity | Raskin 30% | Low Prosperity | Raskin 30% | Low Prosperity | Raskin 30% |
| | (1) | (2) | (3) | (4) | (5) | (6) |
| (A) Flood in last 5 years | 0.039 | 0.008 | 0.452* | 0.384 | 0.258** | 0.236** |
| | (0.069) | (0.094) | (0.264) | (0.240) | (0.103) | (0.107) |
| (B) Flood 0-1 year ago | 0.177 | 0.282 | 0.820* | 1.403*** | 0.539** | 0.622*** |
| | (0.143) | (0.191) | (0.493) | (0.417) | (0.218) | (0.235) |
| Flood > 1 year ago | 0.020 | -0.034 | 0.400 | 0.084 | 0.206* | 0.141 |
| | (0.072) | (0.100) | (0.275) | (0.278) | (0.111) | (0.124) |
| Mean Outcome | 1.43 | 1.40 | 8.08 | 8.09 | 1.48 | 1.47 |
| Observations | 11448 | 7985 | 10954 | 10981 | 5087 | 5096 |

**Note.** – Columns 1, 3, 5 (Low Prosperity) restrict the sample to the panel of adults and children living in an IFLS village whose head identified as being in the bottom three steps of a six-step prosperity ladder in IFLS4. Columns 2, 4, 6 (Raskin 30%) restricts the sample to the panel of individuals living in urban villages where at least 30% of the households received subsidies for the purchase of rice in IFLS4 as part of the Raskin program targeted to the poor. All regressions control for individual fixed effects, wave dummies, individual and household characteristics, and province*wave fixed effects. All models use data from waves 4 and 5 of the IFLS. Robust standard errors in (parentheses).
* $p < 0.1$; ** $p < 0.05$; *** $p < 0.01$.

**Table A10.** Correlation between PODES floods and reporting of floods in IFLS households

| Variables | (1) HHD Flood |
|---|---|
| Age | -0.006 |
|  | (0.006) |
| At least 1 year of high school | -0.029 |
|  | (0.020) |
| Poor Health | -0.003 |
|  | (0.011) |
| CESD | 0.002* |
|  | (0.001) |
| Percentage of villages flooded in district | 0.198* |
|  | (0.118) |
| Percentage of villages flooded in district * Age | 0.001 |
|  | (0.002) |
| Percentage of villages flooded in district * Female | 0.072 |
|  | (0.059) |
| Percentage of villages flooded in district * Secondary Education | 0.082 |
|  | (0.062) |
| Percentage of villages flooded in district * Poor health | 0.050 |
|  | (0.042) |
| Percentage of villages flooded in district * CESD | 0.000 |
|  | (0.004) |
| F-test (interactions) | 0.90 |
| p-value (interactions) | 0.4806 |
| Individual FE | Yes |
| Prov*Year FE | Yes |
| Mean Outcome | 11.64% |
| Observations | 33979 |

**Note.** – The variable percentage of villages flooded was rescaled from 0-100 to a 0-10 scale. All models use data from waves 4 and 5 of the IFLS. Robust standard errors in (parentheses).
* p < 0.1; ** p < 0.05; *** p < 0.01.

**Table A11.** Checking for flood endogeneity among the RISE sample

| Variables | Any Damage from Floods | |
|---|---|---|
| | Adults | Children |
| | (1) | (2) |
| Poor health | -0.014 | -0.047 |
| | (0.040) | (0.043) |
| Number of morbidities | -0.007 | -0.008 |
| | (0.021) | (0.020) |
| CESD-10 Score | -0.004 | - |
| | (0.005) | |
| Poorer Quintile | -0.013 | 0.011 |
| | (0.054) | (0.056) |
| Middle Quintile | -0.042 | -0.015 |
| | (0.062) | (0.059) |
| Richer Quintile | -0.037 | -0.031 |
| | (0.056) | (0.049) |
| Richest Quintile | -0.092 | -0.062 |
| | (0.063) | (0.053) |
| Age | -0.002 | 0.000 |
| | (0.002) | (0.004) |
| Gender: Male | 0.073 | -0.034 |
| | (0.063) | (0.033) |
| Number of children in household | -0.029 | 0.025 |
| | (0.023) | (0.023) |
| Number of people in household | 0.010 | -0.011 |
| | (0.013) | (0.012) |
| Marital Status: Married | 0.105 | 0.156* |
| | (0.078) | (0.093) |
| Marital Status: Other | 0.133 | 0.225* |
| | (0.097) | (0.122) |
| Primary Activity: Employed | -0.011 | -0.015 |
| | (0.046) | (0.044) |
| Has Secondary Schooling or Higher | 0.082** | 0.069* |
| | (0.040) | (0.035) |
| Settlement FE | YES | YES |
| Observations | 489 | 579 |
| R-squared | 0.338 | 0.301 |
| F-test | 0.83 | 0.94 |
| p-value | 0.643 | 0.5119 |

**Note.** – All independent variables were measured at baseline prior to the flood. For the child sample in column 2, the variables marital status, primary activity and level of education correspond to their caregivers. The reference levels are Quintile (Poorest), Gender (Female), Marital Status (Single), Main Activity (Unemployed, seeking job or housewife), Schooling (Less than one year of secondary schooling or no schooling at all). F-test for all controls except settlement FE. Robust standard errors in (parentheses).
* $p<0.1$; ** $p<0.05$; *** $p<0.01$.

**Table A12.** RISE Alternative Specification (Settlement Fixed Effects)

| Variables | Poor Health (1) | Acute Morbidities (2) | Depression Score / Emotional Problems (3) |
|---|---|---|---|
| (A) Adults | | | |
| 4-5 months after | 0.136** | 0.318* | - |
|  | (0.058) | (0.150) |  |
| 10-11 months after | 0.030 | 0.020 | 1.302* |
|  | (0.050) | (0.150) | (0.660) |
| Mean Outcome | 28% | 0.53 | 6.49 |
| Observations | 1429 | 1429 | 897 |
| (B) Children | | | |
| 4-5 months after | 0.109 | 0.195* | 12.325*** |
|  | (0.117) | (0.095) | (2.063) |
| 10-11 months after | -0.019 | -0.064 | 5.864** |
|  | (0.049) | (0.082) | (2.006) |
| Mean Outcome | 17% | 0.59 | 17.38 |
| Observations | 1716 | 1715 | 1030 |

**Note.** - All models control for individual and household characteristics as well as wave fixed effects and settlement*wave interactions. Individual characteristics include: age and gender. Household characteristics include: number of children and number of people in the household. House characteristics include: a list of all assets in the house at baseline, material of floor, roof and walls. All models include data from three waves: Baseline, Wave 2 (4-5 months post flood) and Wave 3 (10-11 months post flood). Robust standard errors in (parentheses).
* $p<0.1$; ** $p<0.05$; *** $p<0.01$.